\documentclass[twocolumn,showpacs,preprintnumbers,amsmath,amssymb,prl, superscriptaddress]{revtex4-1}

\usepackage{lineno}
\usepackage{graphicx}
\usepackage{dcolumn}
\usepackage{bm}
\usepackage{amsmath}
\usepackage{amssymb} 
\usepackage{relsize}
\begin{document}

\title{First observation of PeV-energy neutrinos with IceCube}
\date{\today}

\affiliation{III. Physikalisches Institut, RWTH Aachen University, D-52056 Aachen, Germany}
\affiliation{School of Chemistry \& Physics, University of Adelaide, Adelaide SA, 5005 Australia}
\affiliation{Dept.~of Physics and Astronomy, University of Alaska Anchorage, 3211 Providence Dr., Anchorage, AK 99508, USA}
\affiliation{CTSPS, Clark-Atlanta University, Atlanta, GA 30314, USA}
\affiliation{School of Physics and Center for Relativistic Astrophysics, Georgia Institute of Technology, Atlanta, GA 30332, USA}
\affiliation{Dept.~of Physics, Southern University, Baton Rouge, LA 70813, USA}
\affiliation{Dept.~of Physics, University of California, Berkeley, CA 94720, USA}
\affiliation{Lawrence Berkeley National Laboratory, Berkeley, CA 94720, USA}
\affiliation{Institut f\"ur Physik, Humboldt-Universit\"at zu Berlin, D-12489 Berlin, Germany}
\affiliation{Fakult\"at f\"ur Physik \& Astronomie, Ruhr-Universit\"at Bochum, D-44780 Bochum, Germany}
\affiliation{Physikalisches Institut, Universit\"at Bonn, Nussallee 12, D-53115 Bonn, Germany}
\affiliation{Universit\'e Libre de Bruxelles, Science Faculty CP230, B-1050 Brussels, Belgium}
\affiliation{Vrije Universiteit Brussel, Dienst ELEM, B-1050 Brussels, Belgium}
\affiliation{Dept.~of Physics, Chiba University, Chiba 263-8522, Japan}
\affiliation{Dept.~of Physics and Astronomy, University of Canterbury, Private Bag 4800, Christchurch, New Zealand}
\affiliation{Dept.~of Physics, University of Maryland, College Park, MD 20742, USA}
\affiliation{Dept.~of Physics and Center for Cosmology and Astro-Particle Physics, Ohio State University, Columbus, OH 43210, USA}
\affiliation{Dept.~of Astronomy, Ohio State University, Columbus, OH 43210, USA}
\affiliation{Dept.~of Physics, TU Dortmund University, D-44221 Dortmund, Germany}
\affiliation{Dept.~of Physics, University of Alberta, Edmonton, Alberta, Canada T6G 2E1}
\affiliation{D\'epartement de physique nucl\'eaire et corpusculaire, Universit\'e de Gen\`eve, CH-1211 Gen\`eve, Switzerland}
\affiliation{Dept.~of Physics and Astronomy, University of Gent, B-9000 Gent, Belgium}
\affiliation{Dept.~of Physics and Astronomy, University of California, Irvine, CA 92697, USA}
\affiliation{Laboratory for High Energy Physics, \'Ecole Polytechnique F\'ed\'erale, CH-1015 Lausanne, Switzerland}
\affiliation{Dept.~of Physics and Astronomy, University of Kansas, Lawrence, KS 66045, USA}
\affiliation{Dept.~of Astronomy, University of Wisconsin, Madison, WI 53706, USA}
\affiliation{Dept.~of Physics and Wisconsin IceCube Particle Astrophysics Center, University of Wisconsin, Madison, WI 53706, USA}
\affiliation{Institute of Physics, University of Mainz, Staudinger Weg 7, D-55099 Mainz, Germany}
\affiliation{Universit\'e de Mons, 7000 Mons, Belgium}
\affiliation{T.U. Munich, D-85748 Garching, Germany}
\affiliation{Bartol Research Institute and Department of Physics and Astronomy, University of Delaware, Newark, DE 19716, USA}
\affiliation{Dept.~of Physics, University of Oxford, 1 Keble Road, Oxford OX1 3NP, UK}
\affiliation{Dept.~of Physics, University of Wisconsin, River Falls, WI 54022, USA}
\affiliation{Oskar Klein Centre and Dept.~of Physics, Stockholm University, SE-10691 Stockholm, Sweden}
\affiliation{Department of Physics and Astronomy, Stony Brook University, Stony Brook, NY 11794-3800, USA}
\affiliation{Dept.~of Physics and Astronomy, University of Alabama, Tuscaloosa, AL 35487, USA}
\affiliation{Dept.~of Astronomy and Astrophysics, Pennsylvania State University, University Park, PA 16802, USA}
\affiliation{Dept.~of Physics, Pennsylvania State University, University Park, PA 16802, USA}
\affiliation{Dept.~of Physics and Astronomy, Uppsala University, Box 516, S-75120 Uppsala, Sweden}
\affiliation{Dept.~of Physics, University of Wuppertal, D-42119 Wuppertal, Germany}
\affiliation{DESY, D-15735 Zeuthen, Germany}
          
\author{M.~G.~Aartsen}
\affiliation{School of Chemistry \& Physics, University of Adelaide, Adelaide SA, 5005 Australia}
\author{R.~Abbasi}
\affiliation{Dept.~of Physics and Wisconsin IceCube Particle Astrophysics Center, University of Wisconsin, Madison, WI 53706, USA}
\author{Y.~Abdou}
\affiliation{Dept.~of Physics and Astronomy, University of Gent, B-9000 Gent, Belgium}
\author{M.~Ackermann}
\affiliation{DESY, D-15735 Zeuthen, Germany}
\author{J.~Adams}
\affiliation{Dept.~of Physics and Astronomy, University of Canterbury, Private Bag 4800, Christchurch, New Zealand}
\author{J.~A.~Aguilar}
\affiliation{D\'epartement de physique nucl\'eaire et corpusculaire, Universit\'e de Gen\`eve, CH-1211 Gen\`eve, Switzerland}
\author{M.~Ahlers}
\affiliation{Dept.~of Physics and Wisconsin IceCube Particle Astrophysics Center, University of Wisconsin, Madison, WI 53706, USA}
\author{D.~Altmann}
\affiliation{Institut f\"ur Physik, Humboldt-Universit\"at zu Berlin, D-12489 Berlin, Germany}
\author{J.~Auffenberg}
\affiliation{Dept.~of Physics and Wisconsin IceCube Particle Astrophysics Center, University of Wisconsin, Madison, WI 53706, USA}
\author{X.~Bai}
\thanks{Physics Department, South Dakota School of Mines and Technology, Rapid City, SD 57701, USA}
\affiliation{Bartol Research Institute and Department of Physics and Astronomy, University of Delaware, Newark, DE 19716, USA}
% ...

\collaboration{IceCube Collaboration}\noaffiliation 

\begin{abstract}
We report on the observation of two neutrino-induced events which have an estimated deposited energy in the IceCube detector of $1.04 \pm 0.16$ and $1.14 \pm 0.17$\,PeV, respectively, the highest neutrino energies observed so far. 
These events are consistent with fully contained particle showers induced by neutral-current $\nu_{e,\mu,\tau}$  ($\bar\nu_{e,\mu,\tau}$) or charged-current $\nu_{e}$ ($\bar\nu_{e}$) interactions within the IceCube detector. The events were discovered in a search for ultra-high energy neutrinos using data corresponding to 615.9 days effective livetime. The expected number of atmospheric background is $0.082 \pm 0.004 \text{(stat)}^{+0.041}_{-0.057} \text{(syst)}$. The probability to observe two or more candidate events under the atmospheric background-only hypothesis is $2.9\times10^{-3}$ ($2.8\sigma$) taking into account the uncertainty on the expected number of background events. 
These two events could be a first indication of an astrophysical neutrino flux, the moderate significance, however, does not permit a definitive conclusion at this time.
\end{abstract}

\maketitle

Astrophysical neutrinos are key probes of the high-energy universe. Because of their 
unique properties, neutrinos escape even dense regions, are undeflected in galactic or 
extra-galactic magnetic fields and traverse the photon-filled universe unhindered. 
Thus, neutrinos provide direct information about the dynamics and interiors of cosmological 
objects of the high red-shift Universe like gamma-ray bursts and active galactic nuclei. Neutrinos at 
energies above several 100\,TeV are particularly interesting as the atmospheric background 
in this region is very low and a few astrophysical neutrinos can be significant. 
This letter reports on the observation of two high-energy particle shower events discovered in 
a search for ultra-high energy neutrinos above about 1\,PeV using the IceCube detector.
\begin{figure}[bpt]
  \includegraphics[width=3.2in]{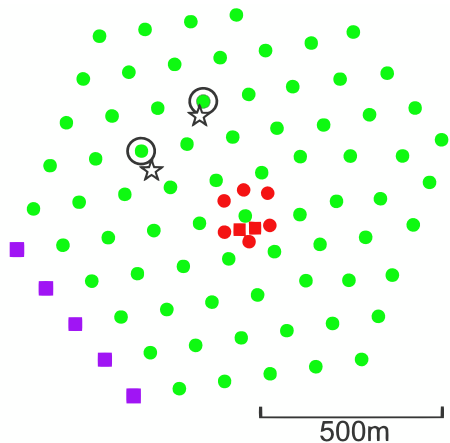}
  \caption{Surface view of the full IceCube detector layout. Filled marks 
    represent the positions of the IceCube strings. Red marks are the DeepCore strings. 
    Squares represent the strings that did not exist in the IC79 configuration. Open circles 
    are the positions of the closest strings to the observed two cascade events.
    Stars are their reconstructed vertex positions. }
    \label{fig:geometry}
\end{figure}

IceCube~\cite{icecube} detects and reconstructs neutrinos by recording Cherenkov photons 
emitted from secondary charged particles produced in neutral-current (NC) or charged-current (CC) interactions of the neutrinos in the 
2,800\,m thick glacial ice at the geographic South Pole. 
IceCube was built between 2005 and 2010.
It consists of an array of 5,160 optical sensors (digital optical modules, DOMs) on 
86 strings at depths between 1,450\,m and 2,450\,m that instrument a volume of 1\,km$^3$ of 
ice. Eight of the 86 strings belong to the DeepCore sub-array~\cite{deepcore}, a more densely 
instrumented volume in the bottom center of the detector. Each DOM consists of a 10" 
photomultiplier tube (PMT)~\cite{pmt} in a spherical glass pressure vessel.
Events are recorded as a series of pulses (waveform) in each DOM~\cite{digitizer} where 
two basic neutrino event signatures are distinguished:
a track-like light pattern originating from neutrino-induced muons (tracks) 
and a spherical light pattern produced by hadronic or electromagnetic particle 
showers (cascades). 

The analysis selects neutrino candidates calorimetrically using the total number of observed photo-electrons in 
each event (NPE)~\cite{digitizer} as a proxy of the deposited energy~\cite{icecubeEHE2010}, thus retaining both bright tracks and cascades.
Backgrounds come from muons and neutrinos generated in interactions of cosmic rays in the 
atmosphere. 
Because of their steeply falling energy spectra, little background is 
expected in the signal region above 1\,PeV. 
The zenith angle distribution of atmospheric muons peaks in the downward-going direction and sharply decreases towards the horizon with a cut-off at a zenith angle $\theta$ of 
cos $\theta$ $\approx 0.15$ due to absorption in the Earth. The atmospheric neutrino distributions have a weaker zenith-angle dependence.
%
The analysis rejects downward-going atmospheric muons by employing event reconstructions based on a track hypothesis in combination with a higher NPE selection criterion in the downward-going region. All remaining events above the combined NPE threshold are considered to be signal candidates independent of their topological properties.
\begin{figure*}[bpt]
  \begin{center}
    \includegraphics[width=2.3in]{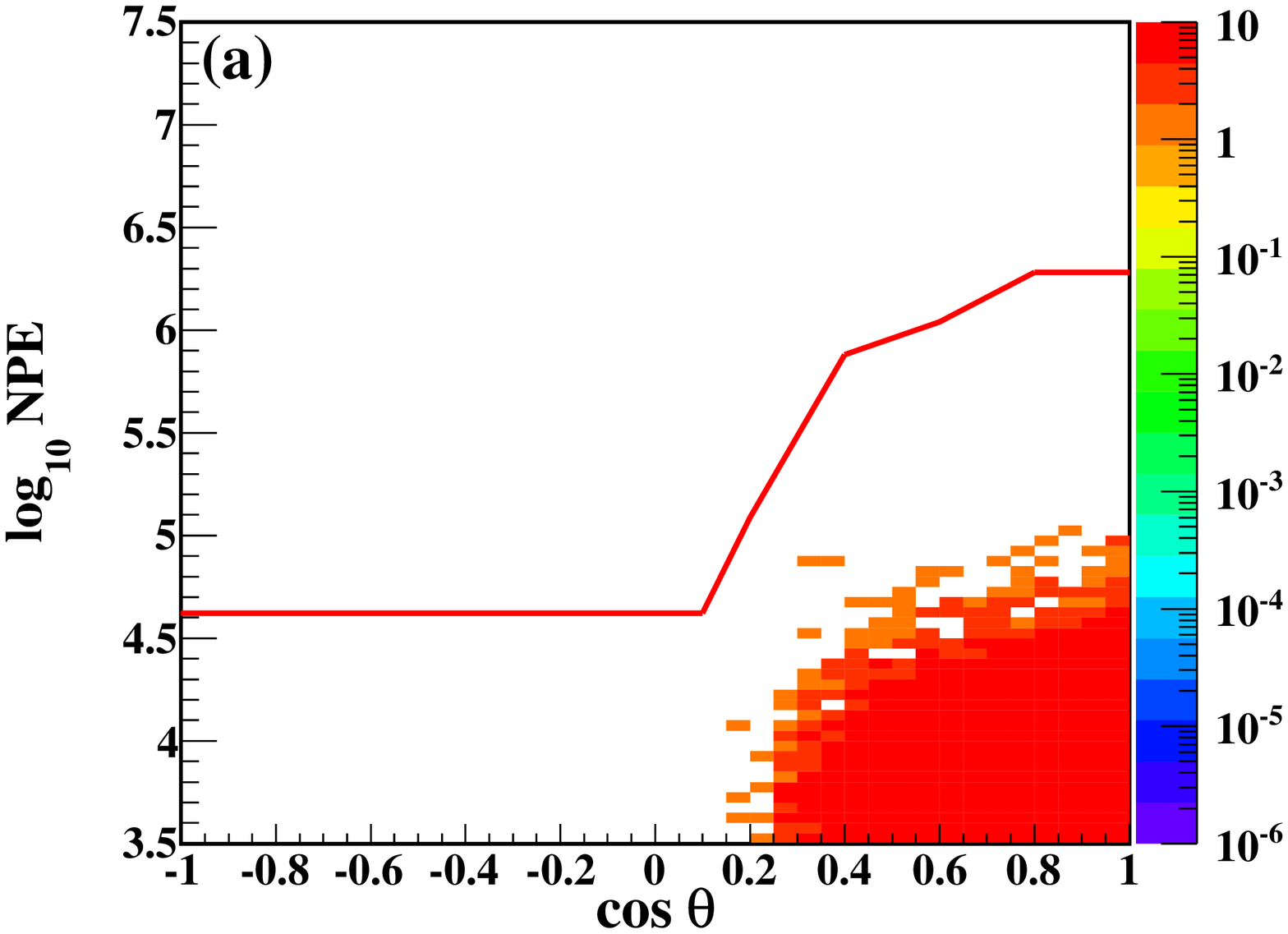}
    \includegraphics[width=2.3in]{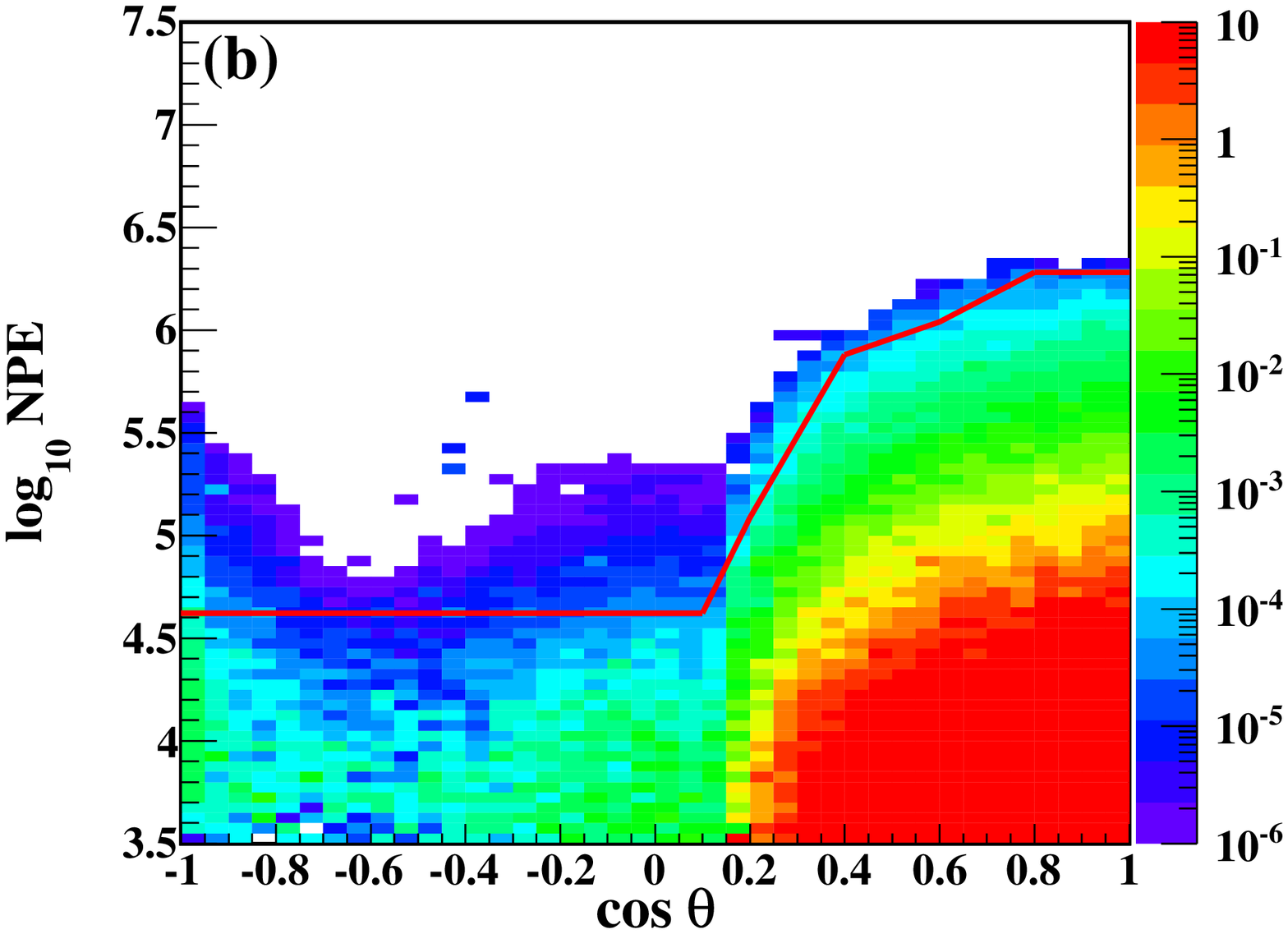}
    \includegraphics[width=2.3in]{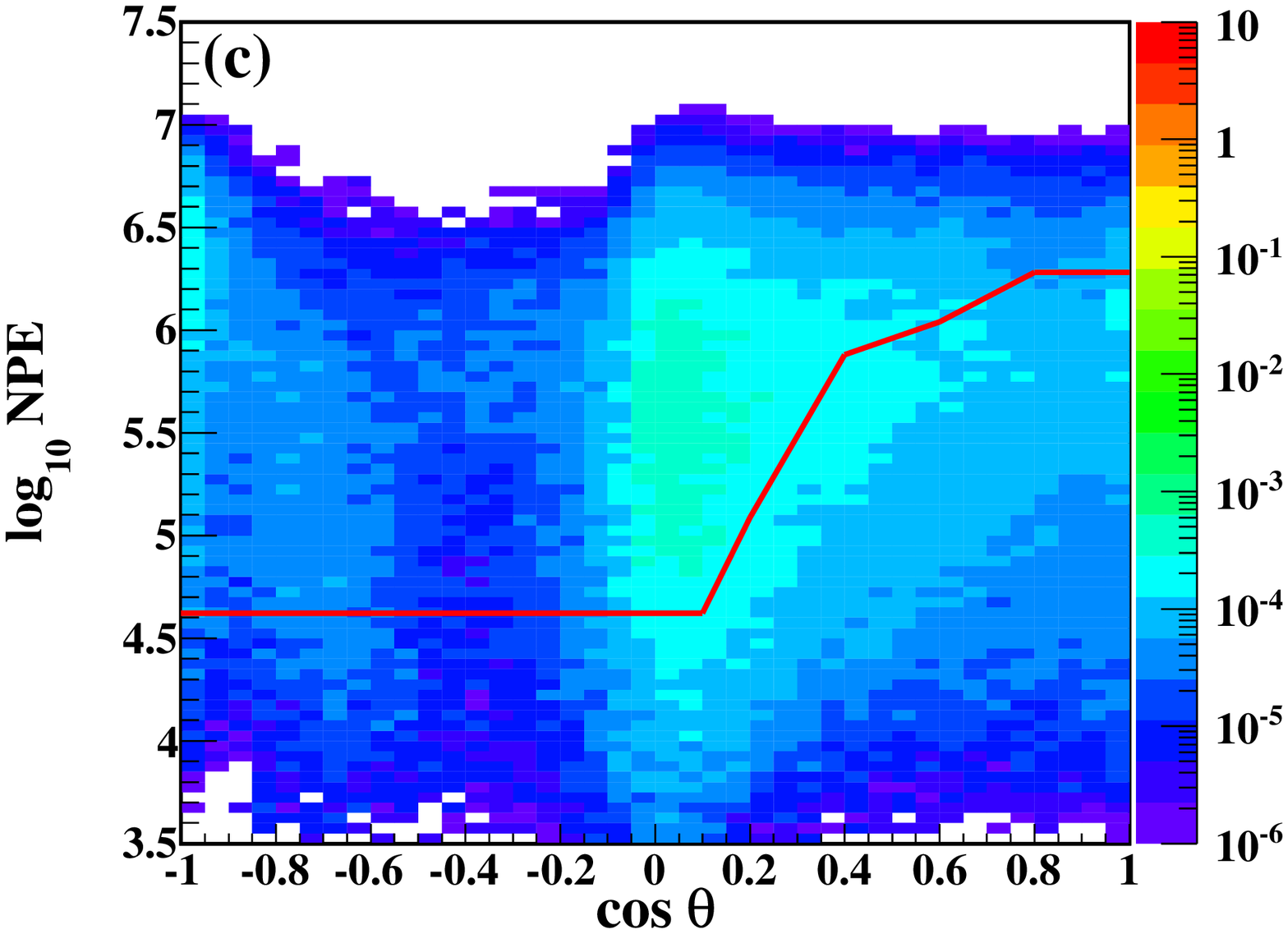}\\
  \end{center}
  \caption{Distribution of NPE and reconstructed zenith angle for (a) the IC79 experimental test sample, (b) the total background, and (c) cosmogenic signal neutrino~\cite{yoshida93}. 
    The colors show event numbers per livetime of 33.4~days.
The solid lines represent the final selection criteria for IC79.}
  \label{fig:final}
\end{figure*}

Data were collected between May 2010 and May 2012, an effective livetime of 615.9\,days 
excluding 54.2\,days used for the optimization of the analysis.
From May 2010 to May 2011, DOMs on 79 strings (IC79) were operational 
(285.8~days livetime with 33.4~days excluded). This period 
was immediately followed by the first year data taking with the full 86-string 
(IC86) detector that lasted until May 2012 (330.1~days livetime with 20.8~days excluded). 
The IC86 configuration is shown in Fig.~\ref{fig:geometry}. 
Events are triggered when eight or more DOMs record signals in local coincidences which occur 
when a nearest or next-to-nearest DOM on the same string triggers within $\pm 1$~$\mu$s~\cite{digitizer}.

The data are filtered at the South Pole with a condition NPE $\geq 1000$, 
and then sent to a northern computer farm via satellite.
In order to avoid biases, we performed a blind analysis and only $\sim$10\% of the data 
were used to develop the analysis.
Photon arrival times are extracted from each waveform and stored as ``hits".
To remove hits from coincident noise, a two-staged cleaning based on the spatial separation 
and the time interval between hits is applied. Data from the DeepCore strings are discarded 
to maintain uniformity across the detector volume. 
To reject downward-going atmospheric muon background, only events with at least 300 hits and 
NPE $\geq 3200$ are retained. To further reduce this background, the directions of the remaining 
events are reconstructed with a track hypothesis and a stricter NPE criterion for downward-going 
tracks is applied (see Fig.~\ref{fig:final} and Eq.~\ref{eqn:selection}):
for IC79, a log-likelihood (LLH) fit is performed \cite{Ahrens2004169} and an event selection 
based on a fit quality parameter is applied to remove events which contain muons from independent 
air showers. 
For IC86, a robust regression technique~\cite{huber, ilf} is utilized to remove hits that have 
a timing significantly different from what is expected from the bulk of the photons from a muon track. 
Afterwards, the direction of the particle is reconstructed with a basic algorithm that assumes a 
plane-wave of photons traveling along the direction of the muon, ``LineFit''~\cite{icecubeEHE2010}.
Both algorithms reconstruct muon tracks with a zenith angle resolution of $1^{\circ}$ or better. 

Cascade events which pass the initial hit and NPE selection criteria are considered 
signal events and therefore should be affected as little as possible by the 
event rejections just described. As they resemble point-like light sources, 
the reconstruction behavior of the two algorithms is indeed quite different 
finding nearly arbitrary zenith angles, albeit with a tendency toward 
upward-going and horizontal directions for the LLH fit and LineFit, 
respectively. Since for these directions the NPE threshold value is 
lower than for downward-going events (see Fig.~\ref{fig:final} and Eq.~\ref{eqn:selection}), such 
events are retained in the final sample even if they would be rejected 
on account of their true direction.

\begin{figure}[t]
  \begin{center}
    \includegraphics[width=3.2in]{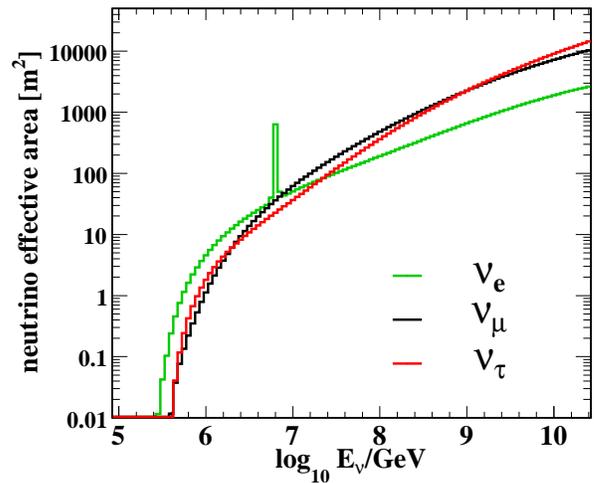}
  \end{center}
  \caption{The average neutrino effective area for a 4$\pi$ isotropic flux, 615.9~days livetime, and the IC79 and IC86 string configurations. Exposure of the sample used in this analysis is obtained by multiplying the effective area with the livetime and 4$\pi$ solid angle. The sharp-peak for $\bar\nu_e$ is the Glashow resonance \cite{glashow}.}
  \label{fig:effarea}
\end{figure}

The NPE threshold values for the two samples were separately optimized based on the simulations to maximize the signal~\cite{icecubeEHE2011, MRF} from the cosmogenic neutrino model \cite{yoshida93}.
Figure~\ref{fig:final} shows the event distributions for the simulations and the experimental IC79 test sample (a livetime of 33.4~days). The solid lines in Fig.~\ref{fig:final} represent the final selection criteria for IC79 where events above the lines constitute the final sample. The final selection criteria for the IC86 sample are:
\begin{equation}
{\rm log_{10} NPE} \geq  \left\{%
	\begin{array}{ll}%
		4.8 & \cos \theta < 0.075 \\
		4.8+1.6  \sqrt{1 - \left(\frac{1.0-\cos\theta}{0.925}\right)^2}&{\cos \theta \geq 0.075.}
	\end{array}\right.
\label{eqn:selection}
\end{equation}
The resulting neutrino effective areas, the equivalent area at the Earth's surface in which neutrinos are detected with 100\% efficiency, averaged over the two-year period from May 2010 to May 2012 taking into account the different detector configurations, is shown in Fig.~\ref{fig:effarea}. The analysis starts to be sensitive in the energy region around $1$\,PeV with its sensitivity rapidly increasing with energy. The effective area is larger for $\nu_e$ than $\nu_\mu$ or $\nu_\tau$ below 10~PeV showing the sensitivity of the present analysis to cascade events in this energy region.

The expected numbers of background events in the final sample for the 615.9~day livetime from atmospheric muons and neutrinos from decays of pions and kaons are 0.038~$\pm0.004$~(stat)~$^{+0.021}_{-0.038}$~(syst) and 0.012~$\pm0.001$~(stat)~$^{+0.010}_{-0.007}$~(syst), respectively.
Compared to previous analyses, the utilized atmospheric neutrino flux models~\cite{TG} accommodate an improved parametrization of the primary cosmic ray spectrum and composition which accounts now for the ``knee'' in the cosmic ray spectrum. Adding prompt atmospheric neutrinos from decays of charmed mesons assuming the model in  \cite{prompt_enberg} with the improved cosmic ray spectrum modeling, the total number of background events increases to 0.082~$\pm0.004$~(stat)~$^{+0.041}_{-0.057}$~(syst). 
Theoretical uncertainties in our {baseline} charmed-meson model~\cite{prompt_enberg} which uses perturbative-QCD calculations are included in the background estimation.
Potential non-perturbative contributions, such as intrinsic charm in nuclei \cite{intrinsic} or from the gluon density at small $x$, could lead to significantly larger cross sections and hence higher prompt neutrino fluxes. Preliminary IceCube limits on the prompt flux at 90\% C.L. are a factor of 3.8 higher than the baseline model~\cite{Anne}.

The main systematic uncertainties on the backgrounds are from the measurement of NPE and from uncertainties in the cosmic ray flux.
They are estimated by varying the associated parameters in the simulation.
The two dominant sources of experimental uncertainties are the absolute DOM sensitivity and the optical properties of the ice which contribute with ($+43$\%, $-26$\%) and ($+0$\%, $-42$\%), respectively. Uncertainties in the cosmic ray flux models are dominated by the primary composition ($+0$\%, $-37$\%) and the flux normalization ($+19$\%, $-26$\%). The theoretical uncertainty in the neutrino production from charm decay~\cite{prompt_enberg} relative to the total background is ($+13$\%, $-16$\%). 
The systematic uncertainties are assumed to be evenly distributed in the estimated allowed range and are summed in quadrature.

%
\begin{figure}[t]
  \begin{center}
    \includegraphics[width=1.67in]{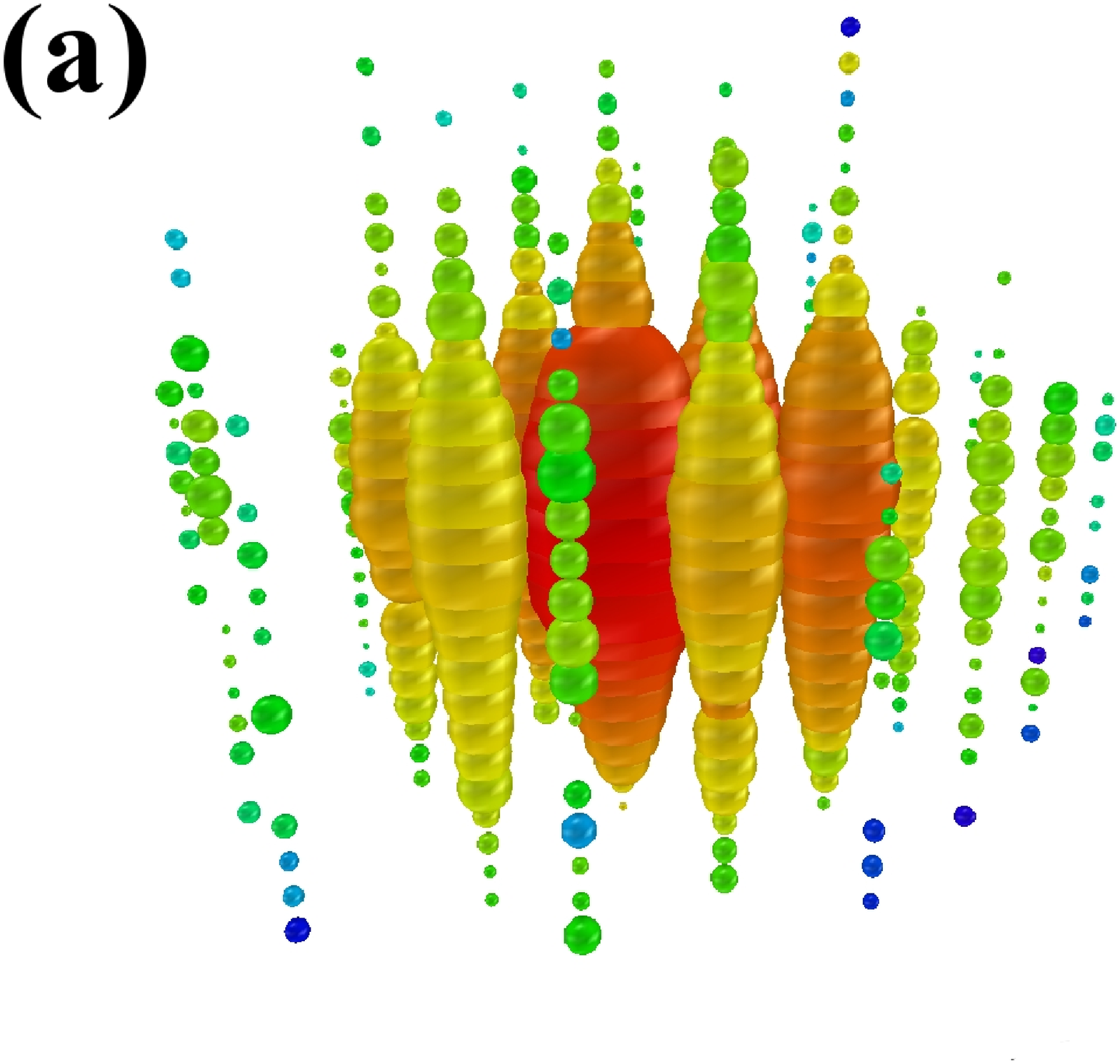}
    \includegraphics[width=1.67in]{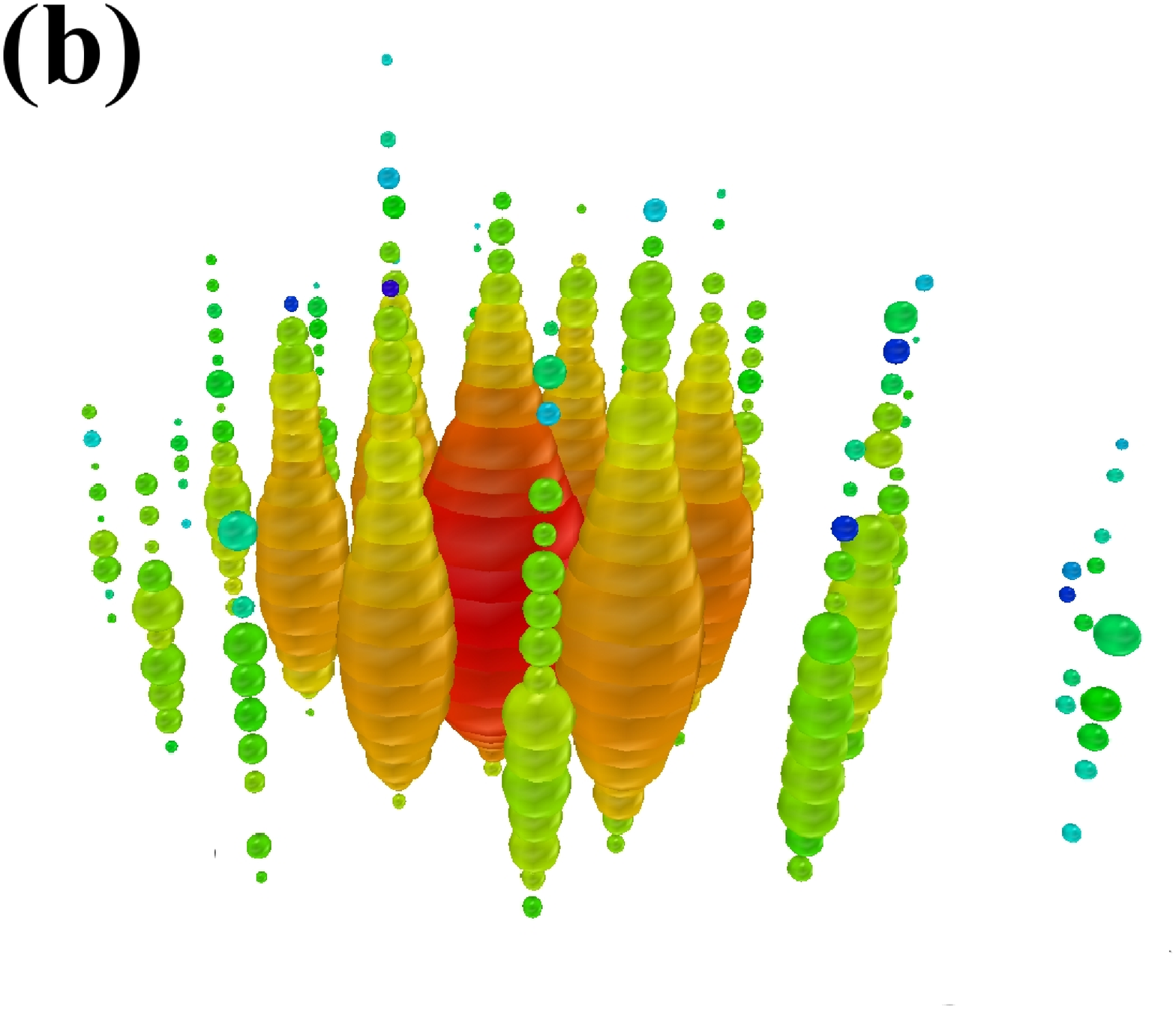}
    \caption{The two observed events from (a) August 2011 and (b) January 2012. Each sphere represents a DOM. Colors represent the arrival times of the photons where red indicates early and blue late times. The size of the spheres is a measure for the recorded number of photo-electrons.}
    \label{fig:eventview}
  \end{center}
\end{figure}

The atmospheric muon and neutrino background events are simulated independently. However, at higher energies, events induced by downward-going atmospheric neutrinos should also contain a significant amount of atmospheric muons produced in the same air shower as the neutrino~\cite{atmnuveto}. 
Since these events are reconstructed as downward-going, they are more likely to be rejected with the higher 
NPE threshold in this region.
Thus, the number of simulated atmospheric neutrino background events is likely overestimated here. 

After unblinding 615.9~days of data, we observe two events that pass all the selection criteria. 
 The hypothesis that the two events are fully explained by atmospheric background including the 
baseline prompt atmospheric neutrino flux~\cite{prompt_enberg} has a p-value of $2.9\times10^{-3}$ ($2.8\sigma$). 
This value includes the uncertainties on the expected number of background events by marginalizing over a flat error distribution.
While the prompt component has 
large theoretical uncertainties, 
obtaining two or more events with a probability of 10\% would require 
a prompt flux that is about 15 times higher than the central value of 
our perturbative-QCD model. This contradicts our preliminary upper limit on 
the prompt flux~\cite{Anne}. Using an extreme prompt
flux at the level of this upper limit which covers a potential unknown 
contribution from intrinsic charm \cite{Thunman1996309}
yields a significance of $2.3\sigma$.

The two events are shown in Fig.~\ref{fig:eventview}. 
\begin{figure}[tbp]
  \begin{center}
    \includegraphics[width=3.2in]{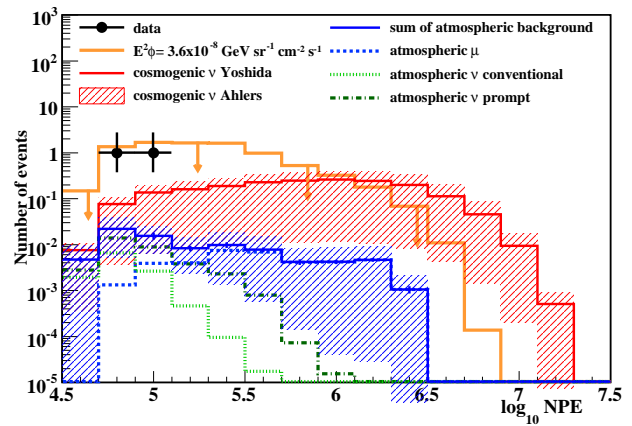}
    \caption{NPE distributions for 615.9~days of livetime at final selection level. 
      The black points are the experimental data. 
      The error bars on the data show the Feldman-Cousins 68\% confidence interval~\cite{FC}.
      The solid blue line marks the sum of the atmospheric muon (dashed blue), conventional atmospheric neutrino (dotted light green) and the baseline prompt atmospheric neutrino (dot-dashed green) background. The error bars on the line and the shaded blue region are the statistical and systematic uncertainties, respectively.
The red line represents the cosmogenic neutrino model \cite{yoshida93}. The shaded region is the allowed level of the cosmogenic $\nu$ flux by Ahlers {\it et al.}~\cite{ahlers2010}.
The orange line represents an $E^{-2}$ power-law flux up to an energy of $10^9$\,GeV with an all-flavor normalization of $E^2 \phi_{\nu_e+\nu_\mu+\nu_\tau}= 3.6\times 10^{-8}\,\text{GeV}\,\text{sr}^{-1}\,\text{s}^{-1}\,\text{cm}^{-2}$, which is the integral upper limit obtained in a previous search in a similar energy range \cite{icecubeEHE2011}. 
The signal fluxes are summed over all neutrino flavors, assuming a flavor ratio of $\nu_e:\nu_\mu:\nu_\tau= 1:1:1$.}
   \label{fig:npe}
  \end{center}
\end{figure}
They are from the IC86 sample, but would have also passed the selection criteria of the IC79 sample.
Their spherical photon distributions are consistent with the pattern of Cherenkov photons from particle cascades induced by neutrino interactions within the IceCube detector. 
There are no indications 
for photons from in-coming or out-going muon or tau tracks. 
Hence, 
these events are most likely induced by either CC interactions of $\nu_e$ or NC interactions of $\nu_e$, $\nu_\mu$ or $\nu_\tau$. 
CC interactions of $\nu_\tau$ induce tau leptons with mean decay lengths of about 50~m at these energies \cite{doublepulse}. The primary neutrino interaction and the secondary tau decay initiate separate cascades which in a fraction of such events lead to an observable double-peak structure in the recorded waveforms. The two events do not show a significant indication of such a signature.
Figure~\ref{fig:npe} shows the final-selection NPE distributions 
for the experimental data, signal models and background simulations. 
The two events are near the NPE threshold of the analysis and are consistent with a 
previous upper limit by IceCube~\cite{icecubeEHE2011} on an unbroken $E^{-2}$  flux, while a flux corresponding to this upper limit predicts about 10 events 
above the NPE cut. The cosmogenic neutrino model~\cite{yoshida93} predicts an event rate of about 2 events
in the corresponding livetime but at significantly higher energies.
%

Maximum-likelihood methods are used to reconstruct the two events.
The likelihood is the product of the Poisson probabilities to observe the recorded number of photo-electrons in a given time interval and DOM for a cascade hypothesis which depends on the interaction vertex, deposited energy and direction. 
Here, the time of the first hit mainly determines the vertex position and the recorded NPE plays a dominant role in estimating the deposited energy.
The hit information used in the reconstruction is extracted from an unfolding procedure of the waveforms.
The open circles in Fig.~\ref{fig:geometry} indicate the strings closest to the reconstructed vertex positions. 
The reconstructed deposited energies of the two cascades are 1.04\,PeV and 1.14\,PeV, respectively, with combined statistical and systematic uncertainties of $\pm$15\% each. 
The errors on the deposited energies are obtained by simulating cascade events in the vicinity of the reconstructed energies and vertices. 
The study is specifically performed on each event and the larger of the two event uncertainties is cited for both events.
Thus, the error associated with the two events differs from that of other cascade events observed in IceCube \cite{ic22cascade}. 
Since there is no absolute energy standard with adequate precision at these energies, the energy scale 
is derived from simulations based on measured ice properties 
and PMT efficiencies which are assured by measurements of atmospheric muons. 
The main sources of systematic uncertainty on the reconstructed deposited energies are the absolute DOM sensitivity and the optical properties of the ice \cite{spice}. The effect of the latter is estimated to be $+9$\% and $-5$\% and is obtained by varying the scattering and absorption coefficients for the photon propagation by 10\%. 
The reconstruction algorithm includes variations of the scattering and absorption coefficients with depth (ice layers) \cite{photonics}.
The effect of a possible azimuthal anisotropy of the ice parameters and a tilt of the ice layers on the reconstructed energies is estimated to be $\pm$5\%. 
The reconstructed energy depends linearly on the DOM efficiency, which has a 10\% systematic 
uncertainty.
The suppression of bremsstrahlung and pair production due to the LPM effect~\cite{LPMSpencer} is negligible in this energy range. 
The properties of the two observed events are summarized in Tab.~\ref{table:theTwoEvents}. 

\begin{table}[tb]
\begin{center}
\begin{tabular}{crr}
\hline
\hline
date (GMT) &  Aug. 8, 2011& Jan. 3, 2012\\
NPE & $7.0 \times 10^4$ & $9.6 \times 10^4$\\
number of recorded DOMs & 354& 312\\
reconstructed deposited  &  &  \\
energy (PeV)& $1.04 \pm 0.16$ & $1.14 \pm 0.17$ \\
reconstructed $z$ vertex (m) &$122 \pm 5$ &$25 \pm 5$\\
\hline
\hline
\end{tabular}
\caption{Characteristics of the two observed events. The depths of the reconstructed vertex positions ``$z$'' are with respect to the center of the IceCube detector at a depth of 1948\,m.}
\label{table:theTwoEvents}
\end{center}
\end{table}

The reconstructed deposited energy is the energy of the incoming neutrino if the observed cascade is the result of a CC interaction of $\nu_e$ neutrino, as in this case the total neutrino energy is deposited near the interaction vertex~\footnotemark. On the other hand, NC interactions of neutrinos of any flavor or interactions of $\overline\nu_e$ via the Glashow resonance at 6.3 PeV~\cite{glashow} with outgoing leptons induce cascades which carry only a fraction of the neutrino energy. The observed cascades are unlikely to originate from the Glashow resonance as only about 10\% of these interactions will deposit 1.2\,PeV or less in the detector in cascade-like signatures.

The two PeV neutrino events observed in two years of data taken with the IceCube neutrino telescope may be a first hint of an astrophysical high-energy neutrino flux. Given the yet rather moderate significance of $2.8 \sigma$ with respect to the expected atmospheric background and the large uncertainties on its prompt component, a firm astrophysical interpretation requires more data in combination with analyses in other detection channels and energy ranges. 

\begin{acknowledgments}

We acknowledge the support from the following agencies:
U.S. National Science Foundation-Office of Polar Programs,
U.S. National Science Foundation-Physics Division,
University of Wisconsin Alumni Research Foundation,
the Grid Laboratory Of Wisconsin (GLOW) grid infrastructure at the University of Wisconsin - Madison, the Open Science Grid (OSG) grid infrastructure;
U.S. Department of Energy, and National Energy Research Scientific Computing Center,
the Louisiana Optical Network Initiative (LONI) grid computing resources;
Natural Sciences and Engineering Research Council of Canada,
Compute Canada and Compute West High Performance Computing;
Swedish Research Council,
Swedish Polar Research Secretariat,
Swedish National Infrastructure for Computing (SNIC),
and Knut and Alice Wallenberg Foundation, Sweden;
German Ministry for Education and Research (BMBF),
Deutsche Forschungsgemeinschaft (DFG),
Helmholtz Alliance for Astroparticle Physics (HAP),
Research Department of Plasmas with Complex Interactions (Bochum), Germany;
Fund for Scientific Research (FNRS-FWO),
FWO Odysseus programme,
Flanders Institute to encourage scientific and technological research in industry (IWT),
Belgian Federal Science Policy Office (Belspo);
University of Oxford, United Kingdom;
Marsden Fund, New Zealand;
Australian Research Council;
Japan Society for Promotion of Science (JSPS);
the Swiss National Science Foundation (SNSF), Switzerland.

\end{acknowledgments}

\begin{thebibliography}{26}%
\makeatletter
\providecommand \@ifxundefined [1]{%
 \@ifx{#1\undefined}
}%
\providecommand \@ifnum [1]{%
 \ifnum #1\expandafter \@firstoftwo
 \else \expandafter \@secondoftwo
 \fi
}%
\providecommand \@ifx [1]{%
 \ifx #1\expandafter \@firstoftwo
 \else \expandafter \@secondoftwo
 \fi
}%
\providecommand \natexlab [1]{#1}%
\providecommand \enquote  [1]{``#1''}%
\providecommand \bibnamefont  [1]{#1}%
\providecommand \bibfnamefont [1]{#1}%
\providecommand \citenamefont [1]{#1}%
\providecommand \href@noop [0]{\@secondoftwo}%
\providecommand \href [0]{\begingroup \@sanitize@url \@href}%
\providecommand \@href[1]{\@@startlink{#1}\@@href}%
\providecommand \@@href[1]{\endgroup#1\@@endlink}%
\providecommand \@sanitize@url [0]{\catcode `\\12\catcode `\$12\catcode
  `\&12\catcode `\#12\catcode `\^12\catcode `\_12\catcode `\%12\relax}%
\providecommand \@@startlink[1]{}%
\providecommand \@@endlink[0]{}%
\providecommand \url  [0]{\begingroup\@sanitize@url \@url }%
\providecommand \@url [1]{\endgroup\@href {#1}{\urlprefix }}%
\providecommand \urlprefix  [0]{URL }%
\providecommand \Eprint [0]{\href }%
\providecommand \doibase [0]{http://dx.doi.org/}%
\providecommand \selectlanguage [0]{\@gobble}%
\providecommand \bibinfo  [0]{\@secondoftwo}%
\providecommand \bibfield  [0]{\@secondoftwo}%
\providecommand \translation [1]{[#1]}%
\providecommand \BibitemOpen [0]{}%
\providecommand \bibitemStop [0]{}%
\providecommand \bibitemNoStop [0]{.\EOS\space}%
\providecommand \EOS [0]{\spacefactor3000\relax}%
\providecommand \BibitemShut  [1]{\csname bibitem#1\endcsname}%
\let\auto@bib@innerbib\@empty
\bibitem [{\citenamefont {Achterberg}\ \emph {et~al.}(2006)\citenamefont
  {Achterberg} \emph {et~al.}}]{icecube}%
  \BibitemOpen
  \bibfield  {author} {\bibinfo {author} {\bibfnamefont {A.}~\bibnamefont
  {Achterberg}} \emph {et~al.} (\bibinfo {collaboration} {IceCube
  Collaboration}),\ }\href {\doibase 10.1016/j.astropartphys.2006.06.007}
  {\bibfield  {journal} {\bibinfo  {journal} {Astropart.Phys.}\ }\textbf
  {\bibinfo {volume} {26}},\ \bibinfo {pages} {155} (\bibinfo {year}
  {2006})}\BibitemShut {NoStop}%
\bibitem [{\citenamefont {Abbasi}\ \emph {et~al.}(2012)\citenamefont {Abbasi}
  \emph {et~al.}}]{deepcore}%
  \BibitemOpen
  \bibfield  {author} {\bibinfo {author} {\bibfnamefont {R.}~\bibnamefont
  {Abbasi}} \emph {et~al.} (\bibinfo {collaboration} {IceCube Collaboration}),\
  }\href {\doibase 10.1016/j.astropartphys.2012.01.004} {\bibfield  {journal}
  {\bibinfo  {journal} {Astropart.Phys.}\ }\textbf {\bibinfo {volume} {35}},\
  \bibinfo {pages} {615} (\bibinfo {year} {2012})}\BibitemShut {NoStop}%
\bibitem [{\citenamefont {Abbasi}\ \emph
  {et~al.}(2010{\natexlab{a}})\citenamefont {Abbasi} \emph {et~al.}}]{pmt}%
  \BibitemOpen
  \bibfield  {author} {\bibinfo {author} {\bibfnamefont {R.}~\bibnamefont
  {Abbasi}} \emph {et~al.} (\bibinfo {collaboration} {IceCube collaboration}),\
  }\href@noop {} {\bibfield  {journal} {\bibinfo  {journal} {Nucl. Instrum.
  Meth.}\ }\textbf {\bibinfo {volume} {{A618}}},\ \bibinfo {pages} {139}
  (\bibinfo {year} {2010}{\natexlab{a}})}\BibitemShut {NoStop}%
\bibitem [{\citenamefont {Abbasi}\ \emph {et~al.}(2009)\citenamefont {Abbasi}
  \emph {et~al.}}]{digitizer}%
  \BibitemOpen
  \bibfield  {author} {\bibinfo {author} {\bibfnamefont {R.}~\bibnamefont
  {Abbasi}} \emph {et~al.} (\bibinfo {collaboration} {IceCube collaboration}),\
  }\href@noop {} {\bibfield  {journal} {\bibinfo  {journal} {Nucl. Instrum.
  Meth.}\ }\textbf {\bibinfo {volume} {A601}},\ \bibinfo {pages} {294}
  (\bibinfo {year} {2009})}\BibitemShut {NoStop}%
\bibitem [{\citenamefont {Abbasi}\ \emph
  {et~al.}(2010{\natexlab{b}})\citenamefont {Abbasi} \emph
  {et~al.}}]{icecubeEHE2010}%
  \BibitemOpen
  \bibfield  {author} {\bibinfo {author} {\bibfnamefont {R.}~\bibnamefont
  {Abbasi}} \emph {et~al.} (\bibinfo {collaboration} {IceCube Collaboration}),\
  }\href {\doibase 10.1103/PhysRevD.82.072003} {\bibfield  {journal} {\bibinfo
  {journal} {Phys.~Rev.}\ }\textbf {\bibinfo {volume} {D82}},\ \bibinfo {pages}
  {072003} (\bibinfo {year} {2010}{\natexlab{b}})}\BibitemShut {NoStop}%
\bibitem [{\citenamefont {Yoshida}\ and\ \citenamefont
  {Teshima}(1993)}]{yoshida93}%
  \BibitemOpen
  \bibfield  {author} {\bibinfo {author} {\bibfnamefont {S.}~\bibnamefont
  {Yoshida}}\ and\ \bibinfo {author} {\bibfnamefont {M.}~\bibnamefont
  {Teshima}},\ }\href {\doibase 10.1143/PTP.89.833} {\bibfield  {journal}
  {\bibinfo  {journal} {Prog.Theor.Phys.}\ }\textbf {\bibinfo {volume} {89}},\
  \bibinfo {pages} {833} (\bibinfo {year} {1993})},\ \bibinfo {note} {the model
  with the source evolution $(z_{max}+1)^m$ with $m = 4$ extending to $z_{max}
  = 4.0$.}\BibitemShut {Stop}%
\bibitem [{\citenamefont {Ahrens}\ \emph {et~al.}(2004)\citenamefont {Ahrens}
  \emph {et~al.}}]{Ahrens2004169}%
  \BibitemOpen
  \bibfield  {author} {\bibinfo {author} {\bibfnamefont {J.}~\bibnamefont
  {Ahrens}} \emph {et~al.},\ }\href {\doibase 10.1016/j.nima.2004.01.065}
  {\bibfield  {journal} {\bibinfo  {journal} {Nucl. Instrum. Meth.}\ }\textbf
  {\bibinfo {volume} {524}},\ \bibinfo {pages} {169 } (\bibinfo {year}
  {2004})}\BibitemShut {NoStop}%
\bibitem [{\citenamefont {Huber}(1981)}]{huber}%
  \BibitemOpen
  \bibfield  {author} {\bibinfo {author} {\bibfnamefont {P.}~\bibnamefont
  {Huber}},\ }\href@noop {} {\emph {\bibinfo {title} {Robust Statistics}}}\
  (\bibinfo  {publisher} {Wiley},\ \bibinfo {address} {New York},\ \bibinfo
  {year} {1981})\BibitemShut {NoStop}%
\bibitem [{\citenamefont {Wellons}\ \emph {et~al.}(2013)\citenamefont {Wellons}
  \emph {et~al.}}]{ilf}%
  \BibitemOpen
  \bibfield  {author} {\bibinfo {author} {\bibfnamefont {M.}~\bibnamefont
  {Wellons}} \emph {et~al.},\ }\href@noop {} {\enquote {\bibinfo {title}
  {Robust statistics in muon reconstruction},}\ } (\bibinfo {year} {2013}),\
  \bibinfo {note} {proceedings of the 33rd International Cosmic Ray Conference
  (ICRC2013), to be published}\BibitemShut {NoStop}%
\bibitem [{\citenamefont {Abbasi}\ \emph
  {et~al.}(2011{\natexlab{a}})\citenamefont {Abbasi} \emph
  {et~al.}}]{icecubeEHE2011}%
  \BibitemOpen
  \bibfield  {author} {\bibinfo {author} {\bibfnamefont {R.}~\bibnamefont
  {Abbasi}} \emph {et~al.} (\bibinfo {collaboration} {IceCube Collaboration}),\
  }\href {\doibase 10.1103/PhysRevD.84.079902, 10.1103/PhysRevD.83.092003}
  {\bibfield  {journal} {\bibinfo  {journal} {Phys.Rev.}\ }\textbf {\bibinfo
  {volume} {D83}},\ \bibinfo {pages} {092003} (\bibinfo {year}
  {2011}{\natexlab{a}})}\BibitemShut {NoStop}%
\bibitem [{\citenamefont {Hill}\ and\ \citenamefont {Rawlins}(2003)}]{MRF}%
  \BibitemOpen
  \bibfield  {author} {\bibinfo {author} {\bibfnamefont {G.}~\bibnamefont
  {Hill}}\ and\ \bibinfo {author} {\bibfnamefont {K.}~\bibnamefont {Rawlins}},\
  }\href@noop {} {\bibfield  {journal} {\bibinfo  {journal} {Astropart.~Phys.}\
  }\textbf {\bibinfo {volume} {19}},\ \bibinfo {pages} {393} (\bibinfo {year}
  {2003})}\BibitemShut {NoStop}%
\bibitem [{\citenamefont {Glashow}(1960)}]{glashow}%
  \BibitemOpen
  \bibfield  {author} {\bibinfo {author} {\bibfnamefont {S.~L.}\ \bibnamefont
  {Glashow}},\ }\href@noop {} {\bibfield  {journal} {\bibinfo  {journal}
  {Physical Review}\ }\textbf {\bibinfo {volume} {118}},\ \bibinfo {pages}
  {316} (\bibinfo {year} {1960})}\BibitemShut {NoStop}%
\bibitem [{\citenamefont {Gaisser}(2012)}]{TG}%
  \BibitemOpen
  \bibfield  {author} {\bibinfo {author} {\bibfnamefont {T.~K.}\ \bibnamefont
  {Gaisser}},\ }\href {\doibase 10.1016/j.astropartphys.2012.02.010} {\bibfield
   {journal} {\bibinfo  {journal} {Astropart.Phys.}\ }\textbf {\bibinfo
  {volume} {35}},\ \bibinfo {pages} {801} (\bibinfo {year} {2012})}\BibitemShut
  {NoStop}%
\bibitem [{\citenamefont {Enberg}\ \emph {et~al.}(2008)\citenamefont {Enberg},
  \citenamefont {Reno},\ and\ \citenamefont {Sarcevic}}]{prompt_enberg}%
  \BibitemOpen
  \bibfield  {author} {\bibinfo {author} {\bibfnamefont {R.}~\bibnamefont
  {Enberg}}, \bibinfo {author} {\bibfnamefont {M.~H.}\ \bibnamefont {Reno}}, \
  and\ \bibinfo {author} {\bibfnamefont {I.}~\bibnamefont {Sarcevic}},\ }\href
  {\doibase 10.1103/PhysRevD.78.043005} {\bibfield  {journal} {\bibinfo
  {journal} {Phys.Rev.}\ }\textbf {\bibinfo {volume} {D78}},\ \bibinfo {pages}
  {043005} (\bibinfo {year} {2008})}\BibitemShut {NoStop}%
\bibitem [{\citenamefont {Brodsky}\ \emph {et~al.}(1980)\citenamefont {Brodsky}
  \emph {et~al.}}]{intrinsic}%
  \BibitemOpen
  \bibfield  {author} {\bibinfo {author} {\bibfnamefont {S.}~\bibnamefont
  {Brodsky}} \emph {et~al.},\ }\href@noop {} {\bibfield  {journal} {\bibinfo
  {journal} {Phys. Lett. B}\ }\textbf {\bibinfo {volume} {93}},\ \bibinfo
  {pages} {451} (\bibinfo {year} {1980})}\BibitemShut {NoStop}%
\bibitem [{\citenamefont {Schukraft}(2013)}]{Anne}%
  \BibitemOpen
  \bibfield  {author} {\bibinfo {author} {\bibfnamefont {A.}~\bibnamefont
  {Schukraft}} (\bibinfo {collaboration} {IceCube Collaboration}),\ }\href@noop
  {} {\enquote {\bibinfo {title} {{A view of prompt atmospheric neutrinos with
  IceCube}},}\ } (\bibinfo {year} {2013}),\ \bibinfo {note}
  {arXiv:1302.0127}\BibitemShut {NoStop}%
\bibitem [{\citenamefont {Sch\"onert}\ \emph {et~al.}(2009)\citenamefont
  {Sch\"onert}, \citenamefont {Gaisser}, \citenamefont {Resconi},\ and\
  \citenamefont {Schulz}}]{atmnuveto}%
  \BibitemOpen
  \bibfield  {author} {\bibinfo {author} {\bibfnamefont {S.}~\bibnamefont
  {Sch\"onert}}, \bibinfo {author} {\bibfnamefont {T.~K.}\ \bibnamefont
  {Gaisser}}, \bibinfo {author} {\bibfnamefont {E.}~\bibnamefont {Resconi}}, \
  and\ \bibinfo {author} {\bibfnamefont {O.}~\bibnamefont {Schulz}},\ }\href
  {\doibase 10.1103/PhysRevD.79.043009} {\bibfield  {journal} {\bibinfo
  {journal} {Phys. Rev. D}\ }\textbf {\bibinfo {volume} {79}},\ \bibinfo
  {pages} {043009} (\bibinfo {year} {2009})}\BibitemShut {NoStop}%
\bibitem [{\citenamefont {Thunman}\ \emph {et~al.}(1996)\citenamefont
  {Thunman}, \citenamefont {Ingelman},\ and\ \citenamefont
  {Gondolo}}]{Thunman1996309}%
  \BibitemOpen
  \bibfield  {author} {\bibinfo {author} {\bibfnamefont {M.}~\bibnamefont
  {Thunman}}, \bibinfo {author} {\bibfnamefont {G.}~\bibnamefont {Ingelman}}, \
  and\ \bibinfo {author} {\bibfnamefont {P.}~\bibnamefont {Gondolo}},\ }\href
  {\doibase 10.1016/0927-6505(96)00033-3} {\bibfield  {journal} {\bibinfo
  {journal} {Astroparticle Physics}\ }\textbf {\bibinfo {volume} {5}},\
  \bibinfo {pages} {309 } (\bibinfo {year} {1996})}\BibitemShut {NoStop}%
\bibitem [{\citenamefont {Feldman}\ and\ \citenamefont {Cousins}(1998)}]{FC}%
  \BibitemOpen
  \bibfield  {author} {\bibinfo {author} {\bibfnamefont {G.~J.}\ \bibnamefont
  {Feldman}}\ and\ \bibinfo {author} {\bibfnamefont {R.~D.}\ \bibnamefont
  {Cousins}},\ }\href {\doibase 10.1103/PhysRevD.57.3873} {\bibfield  {journal}
  {\bibinfo  {journal} {Phys. Rev. D}\ }\textbf {\bibinfo {volume} {57}},\
  \bibinfo {pages} {3873} (\bibinfo {year} {1998})}\BibitemShut {NoStop}%
\bibitem [{\citenamefont {Ahlers}\ \emph {et~al.}(2010)\citenamefont {Ahlers},
  \citenamefont {Anchordoqui}, \citenamefont {Gonzalez-Garcia}, \citenamefont
  {Halzen},\ and\ \citenamefont {Sarkar}}]{ahlers2010}%
  \BibitemOpen
  \bibfield  {author} {\bibinfo {author} {\bibfnamefont {M.}~\bibnamefont
  {Ahlers}}, \bibinfo {author} {\bibfnamefont {L.~A.}\ \bibnamefont
  {Anchordoqui}}, \bibinfo {author} {\bibfnamefont {M.}~\bibnamefont
  {Gonzalez-Garcia}}, \bibinfo {author} {\bibfnamefont {F.}~\bibnamefont
  {Halzen}}, \ and\ \bibinfo {author} {\bibfnamefont {S.}~\bibnamefont
  {Sarkar}},\ }\href {\doibase 10.1016/j.astropartphys.2010.06.003} {\bibfield
  {journal} {\bibinfo  {journal} {Astropart.Phys.}\ }\textbf {\bibinfo {volume}
  {34}},\ \bibinfo {pages} {106} (\bibinfo {year} {2010})}\BibitemShut
  {NoStop}%
\bibitem [{\citenamefont {Cowen}(2007)}]{doublepulse}%
  \BibitemOpen
  \bibfield  {author} {\bibinfo {author} {\bibfnamefont {D.}~\bibnamefont
  {Cowen}} (\bibinfo {collaboration} {IceCube Collaboration}),\ }\href
  {\doibase 10.1088/1742-6596/60/1/048} {\bibfield  {journal} {\bibinfo
  {journal} {J.Phys.Conf.Ser.}\ }\textbf {\bibinfo {volume} {60}},\ \bibinfo
  {pages} {227} (\bibinfo {year} {2007})}\BibitemShut {NoStop}%
\bibitem [{\citenamefont {Abbasi}\ \emph
  {et~al.}(2011{\natexlab{b}})\citenamefont {Abbasi} \emph
  {et~al.}}]{ic22cascade}%
  \BibitemOpen
  \bibfield  {author} {\bibinfo {author} {\bibfnamefont {R.}~\bibnamefont
  {Abbasi}} \emph {et~al.} (\bibinfo {collaboration} {IceCube Collaboration}),\
  }\href {\doibase 10.1103/PhysRevD.84.072001} {\bibfield  {journal} {\bibinfo
  {journal} {Phys. Rev. D}\ }\textbf {\bibinfo {volume} {84}},\ \bibinfo
  {pages} {072001} (\bibinfo {year} {2011}{\natexlab{b}})}\BibitemShut
  {NoStop}%
\bibitem [{\citenamefont {Aartsen}\ \emph {et~al.}(2013)\citenamefont {Aartsen}
  \emph {et~al.}}]{spice}%
  \BibitemOpen
  \bibfield  {author} {\bibinfo {author} {\bibfnamefont {M.}~\bibnamefont
  {Aartsen}} \emph {et~al.},\ }\href {\doibase 10.1016/j.nima.2013.01.054}
  {\bibfield  {journal} {\bibinfo  {journal} {Nucl. Instrum. Meth.}\ }\textbf
  {\bibinfo {volume} {711}},\ \bibinfo {pages} {73 } (\bibinfo {year}
  {2013})}\BibitemShut {NoStop}%
\bibitem [{\citenamefont {Lundberg}\ \emph {et~al.}(2007)\citenamefont
  {Lundberg} \emph {et~al.}}]{photonics}%
  \BibitemOpen
  \bibfield  {author} {\bibinfo {author} {\bibfnamefont {J.}~\bibnamefont
  {Lundberg}} \emph {et~al.},\ }\href {\doibase 10.1016/j.nima.2007.07.143}
  {\bibfield  {journal} {\bibinfo  {journal} {Nucl. Instrum. Meth.}\ }\textbf
  {\bibinfo {volume} {A581}},\ \bibinfo {pages} {619 } (\bibinfo {year}
  {2007})}\BibitemShut {NoStop}%
\bibitem [{\citenamefont {Gerhardt}\ and\ \citenamefont
  {Klein}(2010)}]{LPMSpencer}%
  \BibitemOpen
  \bibfield  {author} {\bibinfo {author} {\bibfnamefont {L.}~\bibnamefont
  {Gerhardt}}\ and\ \bibinfo {author} {\bibfnamefont {S.~R.}\ \bibnamefont
  {Klein}},\ }\href {\doibase 10.1103/PhysRevD.82.074017} {\bibfield  {journal}
  {\bibinfo  {journal} {Phys. Rev. D}\ }\textbf {\bibinfo {volume} {82}},\
  \bibinfo {pages} {074017} (\bibinfo {year} {2010})}\BibitemShut {NoStop}%
\bibitem [{Note1()}]{Note1}%
  \BibitemOpen
  \bibinfo {note} {The energy reconstruction assumes that all light emission
  originates from an electromagnetic shower. A hadronic cascade with the same
  light yield as the observed events would on average have an about 10\% higher
  energy.}\BibitemShut {Stop}%
\end{thebibliography}
%

               %
\end{document}